\documentclass[a4]{article}
\usepackage[dvips]{graphicx}
\begin{document}
\title{Superluminal Signal Velocity}
\author{G\"unter Nimtz\\
  II.Physikalisches Institut, Universit\"at K\"oln}
\date{\today}
\maketitle
\begin{abstract}It recently has been demonstrated that signals conveyed by evanescent 
modes can travel faster than light. In this report some special features 
of signals are introduced and investigated, for instance the fundamental property that 
signals are frequency band limited. 

Evanescent modes are 
characterized by extraordinary properties: Their energy is 
{\it negative}, they are not directly measurable, and the evanescent 
region is not causal since the modes traverse this region instantaneously. 
The study demonstrates the necessity of quantum mechanical principles in 
order to interprete the superluminal signal velocity of classical 
evanescent modes.

\end{abstract}

\section{Introduction}Tunneling represents the wave mechanical analogy to the propagation of 
evanescent modes \cite{Sommerfeld}. Evanescent modes are observed, e.g. in 
the case of total reflection, in undersized waveguides, and in periodic 
dielectric heterostructures \cite{Nimtz1}. Compared with the 
wave solutions an evanescent mode is characterized by a purely ${\it  
imaginary}$ wave number, so that i.e. the wave equation yields 
for the electric field $E(x)$

\begin{eqnarray}
E(x) =  e^{i(\omega t - k x)} \,\,\,\Rightarrow \,\,\,E(x) = e^{i \omega t 
- \kappa x} ,
\end{eqnarray}
where $\omega$ is the angular frequency, t the time, x the distance, k the 
wave number, and $\kappa = i k$ the ${\it imaginary}$ wave number of the 
evanescent mode.

\begin{figure}
  \begin{minipage}[c]{0.45\linewidth}
   
    \includegraphics[width=\linewidth]{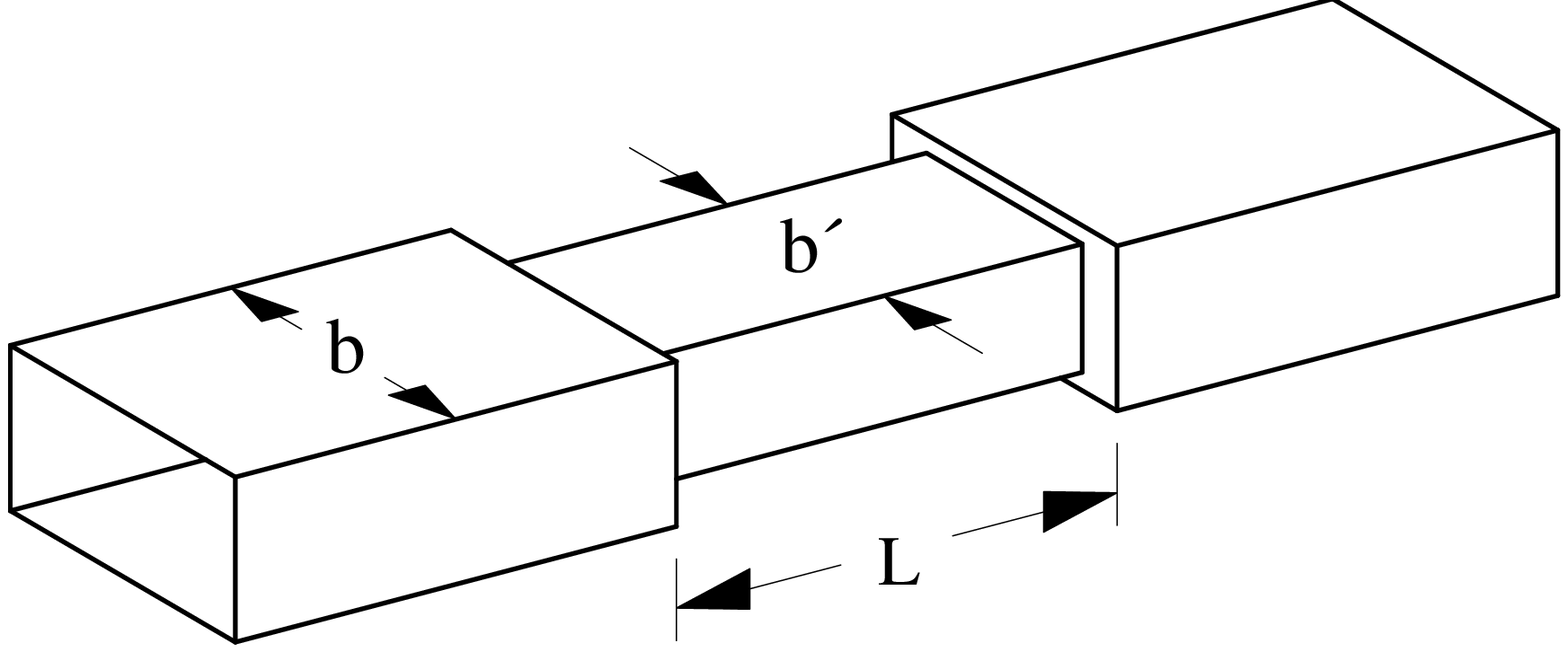}
    \begin{center}
      (a)
    \end{center}
   
    \includegraphics[width=\linewidth]{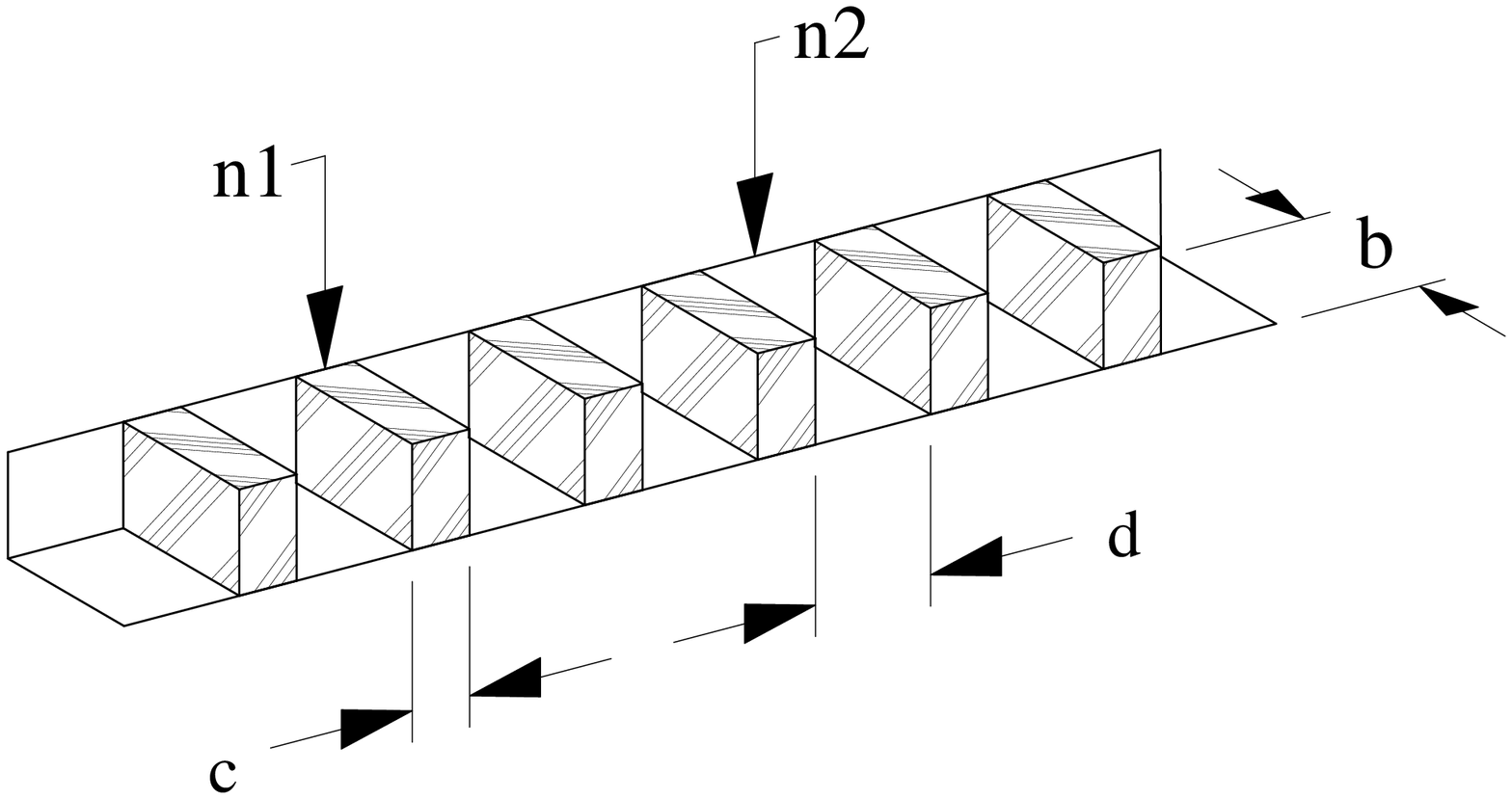}
    \begin{center}
      (b)
    \end{center}
  \end{minipage}
  \hfill
  \begin{minipage}[c]{0.50\linewidth}
    \includegraphics[width=\linewidth]{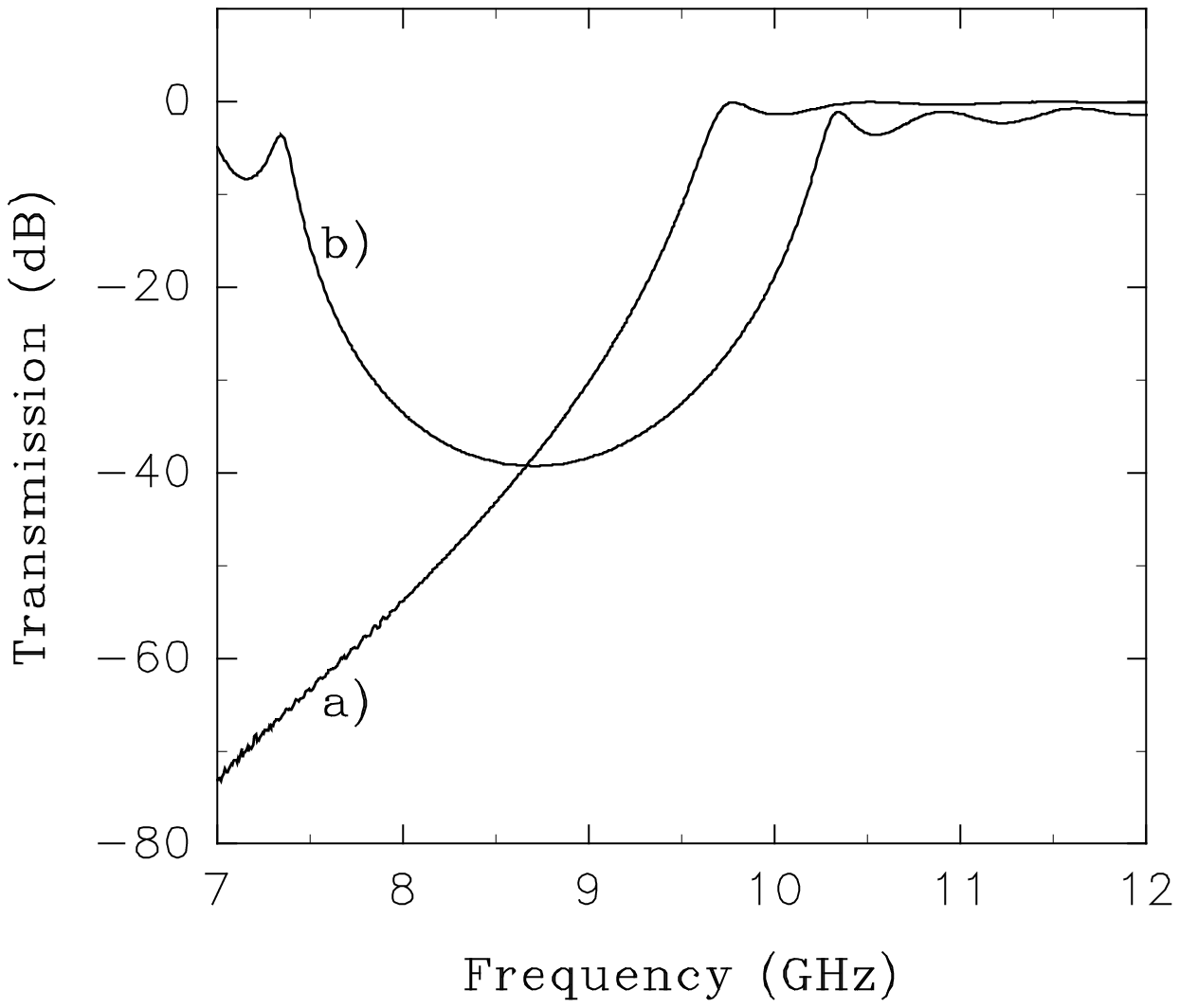}
    \begin{center}
      (c)
    \end{center}
  \end{minipage}
\caption[]{Two examples of electromagnetic structures with evanescent mode solutions: (a) of a waveguide with an undersized 
central part and (b) a one-dimensional periodic hetero-structure. n1 and n2 are the refractive index and c and d the thickness of the dielectric materials. In (c) 
the graphs show the dispersion relations for both structures. The 
transmission dispersion of the periodic heterostructure 
displays a forbidden gap which corresponds to a tunneling regime, 
for details see 
Ref.\cite{Nimtz1}. 
The evanescent regime is characterized by a strong 
attenuation due to the exponential decay.}
\label{Examples}
\end{figure}

Thus evanescent modes are characterized by an exponential attenuation 
and a lack of phase shift. The latter means that the mode has not spent 
time in the evanescent region, which in turn results in an infinite 
velocity in the phase time approximation neglecting the phase shift at the 
boundary \cite{Hartman}. Two examples of electromagnetic structures in 
which evanescent modes exist are shown in Fig.\ref{Examples} 
\cite{Nimtz1}. The dispersion relations of the respective transmission 
coefficients are displayed in the same figure.

\section{Signals}

\begin{figure}
    \includegraphics[width=\linewidth]{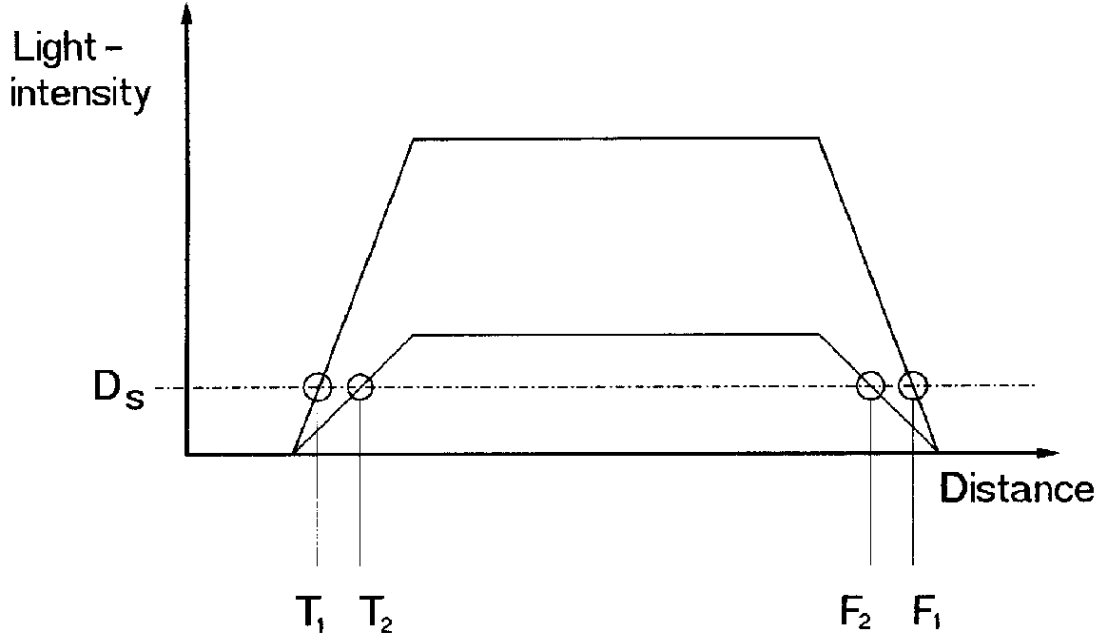}
\caption{Sketch of a light train. The two detected fronts of the same 
attenuated and not attenuated light beam are $F_1$, $F_2$ and the tails are 
$T_1$, $T_2$. They are measured at different places and times in 
consequence of the detector's responsitivity level $D_s$.}
\label{Zeit}
\end{figure}

Quite often a signal is said to be defined by switching on or off 
light. It is assumed that the front of the light beam informs 
my neighbour of my arrival home
with the 
speed of light. 
Such a signal is sketched in Fig.\ref{Zeit}. The inevitable inertia of the 
light source causes an inclination of the signal's front and tail. Due to 
the detector's sensitivity, level $D_S$, the information about the 
neighbours arrival (switching on) and departure (switching off) becomes 
dependent on intensity. In this example the departure time is 
detected earlier with the attenuated weak signal.

This special signal does not transmit reliable information on arrival and 
departure time. In addition to the dependence on intensity , a light front 
may be generated by any spontaneous emission process or accidentally 
by another neighbour. Obviously, a detector needs more than a signal's 
front to response properly. The detector needs information about carrier 
frequency and modulation of the signal in order to obtain reliable 
information about the cause. 

In general, a signal 
is detected some time after the arrival of the light's front. 
Due to the dynamics of detecting systems there are several 
signal oscillations needed in order to produce  
an effect \cite{Sommerfeld}. An ${\it effect}$ is detected with the energy 
velocity. In vacuum or in a medium with normal dispersion the signal 
velocity equals both, the energy and the group velocities. 

\begin{figure}
  \includegraphics[angle=270,width=\linewidth]{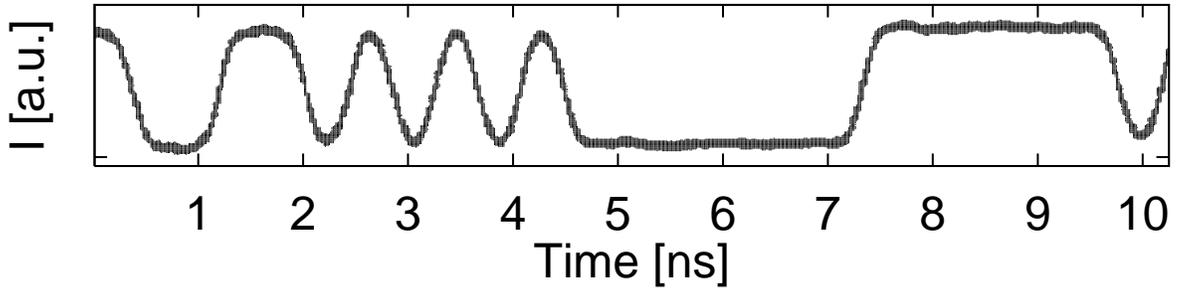}
  \\[3mm]
\caption[]{An example of a modern AM signal used in optoelectronics.
The halfwidth of the pulse-like signal represents the number of digits, 
i.e. 
to the transmitted information.  
The carrier wave frequency is 2 $10^{14}$~Hz and the amplitude modulation 
is limited to a frequency band width of about $10^{10}~$Hz.}
\label{signal}
\end{figure}

For example, a classical signal can be transmitted by the Morse 
alphabet, in which each letter corresponds to a certain number of dots and 
dashes. In general signals are either frequency (FM) or amplitude modulated 
(AM) and they have in common that the signal does not depend on its 
magnitude. A modern signal transmission, where the halfwidth corresponds to 
the number of digits, is displayed in Fig.\ref{signal}. This AM signal has 
an infra-red carrier with $1.5\mu m$ wave length and is glass fiber guided 
from transmitter to receiver.  As mentioned above, the signals are 
independent of magnitude as the halfwidth does not depend on the signal's 
magnitude.

The front or very beginning of a signal  is only well defined 
in the theoretical case of an infinite frequency spectrum. However, 
physical generators only produce signals of finite spectra. This is 
due to their inherent inertia and due to a signal's finite energy content.
These properties result in a real front which is defined by the measurable 
beginning of the signal. For example the signals of Fig.\ref{signal} have a 
detectable frequency band width of $\Delta \nu = \pm 10^{-4} \cdot 
\nu_C$, where $\nu_C$ is the carrier frequency. 

Frequency band limitation in consequence of a finite signal energy 
reveals one of the fundamental deficiencies of classical physics. A 
classical detector can detect a deliberately small amount of energy, 
whereas every physical detector needs at least one quantum of the energy 
$\hbar\,\omega$ in order to respond.

\section{An Experimental Result}
\begin{figure}
\includegraphics[width=\linewidth,clip=]{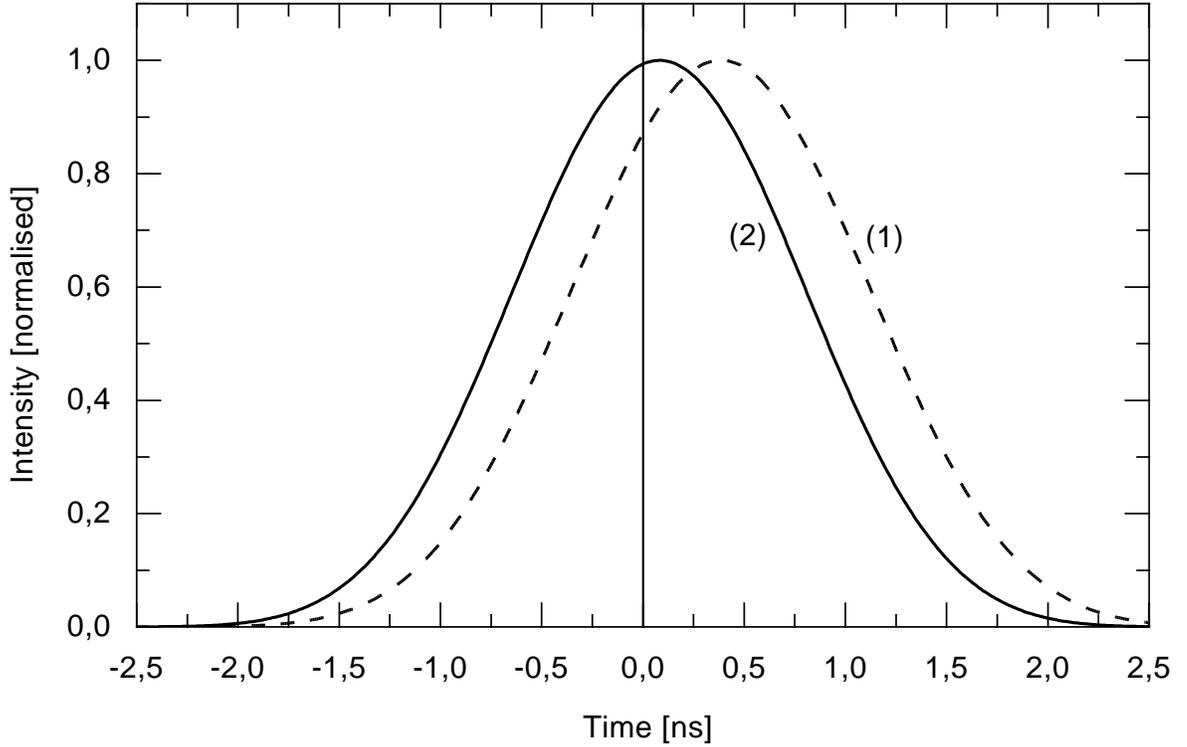}
\caption[]{Measured barrier traversal time of a microwave packet through 
a multilayer structure inside a waveguide (barrier length 114.2~mm). The 
center frequency has been 8.7~GHz with a frequency width of $\pm$ 
0.5~GHz. The pulse's magnitudes are normalized. 
The slow pulse (1) traversed the empty waveguide, whereas the fast one (2) 
has tunneled the forbidden gap of the same length The maximum corresponds  
to the center of mass and equals the group velocity. The group 
velocity of the tunneled signal was 4.7 $\cdot$ c \cite{Nimtz2}. The transmission 
dispersion of the barrier is shown in Fig.1(c) curve b). The tunneled signal (i.e. 
the halfwidth of the pulse) traversed the 114.2~mm long barrier in 81~ps, 
whereas the signal spent 380~ps to cross the same air 
distance. The time resolution in the experiment 
has been better than $\pm$ 1~ps 
\cite{Enders,Nimtz2}.}
\label{Impuls}
\end{figure}

Superluminal signal velocities have been measured by Enders and 
Nimtz \cite{Enders,Enders2,Nimtz2}. The experiments were carried out with 
AM microwaves in undersized waveguides and in periodic 
dielectric heterostructures. The measured propagation time 
of a pulse is shown in Fig.\ref{Impuls}. The microwave pulse has 
travelled either through air or it has crossed an evanescent barrier 
\cite{Nimtz2}. The linewidth of the pulse represents the signal. The 
experimental result is, that the tunneled signal has passed the 
airborne signal at a superluminal velocity of 4.7 $\cdot$ c.
The measurements of the traversal time are carried out under vacuum-like 
conditions at the exit of the evanecscent region, the reason for this will 
be discussed later.

\section{Some Implications of superluminal signal velocity}

Measured microwave signals are shown in Fig.\ref{Impuls}.
The halfwidth (information) of the tunneled signal has traversed the 
evanescent region at a velocity of 4.7 $\cdot$ c.
As explained above, signals have a limited frequency 
spectrum since their energy content W is always finite and 
detectable frequency components with $\omega \ge W/\hbar$ can not exist. 

In this experiment all frequency components of the signal are 
evanescent and move at a velocity faster than c. The beginning of 
the evanescent signal overtakes that of the airborne signal 
as seen in Fig.\ref{Impuls}. The 
superluminal velocity of evanescent modes has some interesting features
differing fundamentally from luminal or subluminal propagation of 
waves with ${\it real}$ wave numbers. This will be discussed in the 
following subsections.

\subsection{Change of chronological order}

The existence of a superluminal signal velocity ensures the posssibility 
of an interchange of chronological order. Let us assume an inertial 
system $\Sigma_{II}$ moves away from system $\Sigma_I$ 
with a velocity $v_r$. The Special 
Relativity (SR) gives the following relationship for the travelling time 
$\Delta t$ and for the distance $\Delta x$ of a signal in 
the system $\Sigma_I$ which is watched in $\Sigma_{II}$

\begin{eqnarray}
\Delta t_{II} = \frac{\Delta_I - v_r \Delta x_I/c^2}{(1 - v_r^2/c^2)^{1/2}}
 = \frac{\Delta t_I(1 - v_S v_r/c^2)}{(1 - v_r^2/c^2)^{1/2}} \qquad.
\end{eqnarray}

$v_r \ge c^2/v_S$ is the condition 
for the change of chronological order, i.e. $\Delta t_{II} \le 
0$, between the systems $\Sigma_I$ and 
$\Sigma_{II}$. For example, at a signal velocity $v_S \ge$ 10 c the 
chronological order changes at $v_r \le$ 0.1 c. This result does not 
violate the SR. The common constraint $v_S$$\le c$ is posed by the SR on 
electromagnetic wave propagation in a dispersive medium and not on the 
propagation of evanescent modes.

\subsection{Negative electromagnetic energy}

The Schr\"odinger equation yields a negative kinetic energy in the 
tunneling case, where the potential U is larger than the particle's 
total energy W:

\begin{eqnarray}
\frac{d^2 \Psi}{d^2x} + \frac{2 m}{\hbar^2} (W - U) \Psi = 0
\end{eqnarray}

The same happens to evanescent modes. Within the mathematical analogy their 
kinetic electromagnetic energy is negative too. The Helmholtz equation for 
the electric field $E$ in a waveguide is given by the relationship

\begin{eqnarray}
\frac{d^2 E}{d^2x} + ( k^2 - k_c^2) E = 0
\end{eqnarray}
where $k_c$ is the cut-off wave number of 
the evanescent regime. The quantity ($k^2$ - $k_c^2$) plays a role 
analogous to the energy eigenvalue and is negative in the case of 
evanescent modes.

The 
dielectric function $\epsilon$ of evanescent modes is negative and thus the 
refractive index is imaginary.

For the basic mode 
a rectangular waveguide
has the 
following dispersion of its dielectric function,
where $k_c$ = 
$(\pi/b)^2$ holds and $b$ is the waveguide width,

\begin{eqnarray}
\epsilon(\lambda_0) = (1 - \lambda_0/2 b)
\end{eqnarray}
\noindent
$\lambda_0$ is the free space wavelength of the electromagnetic wave.

In the case of tunneling it is argued that a particle can only be 
measured in the barrier with a photon having an energy $\hbar \omega \ge$ 
(U - W) \cite{QM}. This means that the total energy of the system is 
positive. According to Eq.(5) the evanescent 
mode's electric energy density $\rho$ is given by the relationship

\begin{eqnarray}
\rho = \frac{1}{2} \,\,\epsilon \,\,\epsilon_0 \,\,\,E^2 \,\,\,\,  {\bf < 
\,0},
\end{eqnarray}

\noindent
where $\epsilon_0$ is the electric permeability.

The analogy between the 
Schr\"odinger equation and the Helmholtz equation holds again and it is not 
possible to measure an evansecent mode.
Achim Enders and I tried hard to measure evanescent modes with 
probes put into the evanescent region but failed 
\cite{Enders3}. Obviously evanescent modes are not directly measurable in 
analogy to a particle in a tunnel. We might also say this problem is due to 
impedance mismatch between the evanescent mode and a probe. The impedance Z 
of the basic mode in a rectangular waveguide is given by the relationship

\begin{eqnarray}
Z = Z_0 \,\, \epsilon^{-1/2} 
\end{eqnarray}
where $Z_0$ is the free space impedance. In the evanesecent regime $k < 
k_c$ the impedance is imaginary.

\subsection{The not-causal evanescent region}

Evanescent modes do not experience a phase shift inside the evanescent 
region \cite{Nimtz1,Hartman}. They cross this region 
without consuming time. The predicted \cite{Hartman} and the measured 
\cite{Nimtz1} time delay happens at the boundary between the wave and the 
evanescent mode regime. For opaque barriers (i.e. $\kappa \cdot x \ge 
1$,where $\kappa$ is the $\it imaginary$ wave number and $x$ the length of 
the evanescent barrier) the phase shift becomes constant with $\approx 2 
\pi$ which corresponds to one oscillation time of the mode. In 
fact, the measured barrier traversal time was roughly equal to the 
reciprocal frequency in the microwave as well as in the optical 
experiments, i.e. either in the 100 ps or in the 2 fs time range 
independent of the barrier length \cite{Nimtz1}. The latter behaviour is 
called {\it Hartman effect}: the tunneling time is independent of barrier 
length and has indeed been measured with 
microwave pulses thirty years after its prediction \cite{Enders2}.

\section{Summing up}

Evanescent modes show some amazing properties with which we are not 
familiar. For instance the evansecent region is not causal since 
evanescent modes do not spend time there. This is an experimental result 
due to the fact that the traversal time is independent of 
barrier length. 

Another strange 
experience in classical physics is that evanescent fields cannot be 
measured. This is due to their ${\it negative}$ 
energy or to the impedance mismatch. Amazingly enough this is in analogy 
with the wave mechanical tunneling.

The energy of signals is always finite resulting in a limited frequency 
spectrum both according to Planck's energy quantum $\hbar \omega$. This 
is a fundamental deficiency of classical physics, which assumes the 
measurability of any small amount of energy. A physical signal never has an ideal front. the 
latter needs infinite high frequency components with a correspondingly 
high energy.

Another consequence of the frequency band limitation of signals is, if they have only 
evanescent mode components, they are not Lorentz-invariant. The signal may 
travel faster than light.

Front, group, signal, and energy velocities all 
have the same value in vacuum. Bearing in mind the narrow frequency band of 
signals, the former statement holds also for the velocities of evanescent 
modes. In first order approximation the dispersion relation of a stop band 
is constant and a significant pulse reshaping does not take place. This 
result demonstrate that signals and effects may transmitted with 
superluminal velocities provided that they are carried by evanescent modes.

\section{Acknowledgments}
Stimulating discussions with V. Grunow, D. Kreimer, P. Mittelstaedt,  R. 
Pelster, and H. Toyatt are gratefully acknowledged.

\end{document}